\documentclass{PoS}

\usepackage{url}            
\usepackage{amsmath}
\usepackage{bm}

\title{
\ \\[-3cm] {\hfill \normalsize \raggedright JLAB-THY-15-2215} \\[2.7cm]
PDFs from nucleons to nuclei
}

\ShortTitle{PDFs from nucleons to nuclei}

\author{\speaker{A.~Accardi}\\
        Hampton U. and Jefferson Lab\\
        E-mail: \email{accardi@jlab.org}}

\abstract{I review recent progress in the extraction of unpolarized parton distributions in the proton and in nuclei from a unified point of view that highlights how the interplay between high energy particle physics and lower energy nuclear physics can be of mutual benefit to either field. Areas of overlap range from the search for physics beyond the standard model at the LHC, to the study of the non perturbative structure of nucleons and the emergence of nuclei from quark and gluon degrees of freedom, to the interaction of colored probes in a cold nuclear medium.
}

\FullConference{
Based on a plenary talk delivered at the\\
XXIII International Workshop on Deep-Inelastic Scattering\\
27 April - May 1 2015\\
Dallas, Texas}

\begin{document}

\section{Introduction}
  
Aside from their importance in calculating high-energy scattering processes of strongly interacting particles, Parton Distribution Functions (PDFs) play an important role in connecting the fields of high-energy particle physics and of lower energy hadronic and nuclear physics \cite{Accardi:2013pra}. This connection is made possible by the universality of PDFs of the proton, which allows one to calculate a variety of processes from a common set of quark and gluon momentum distributions. These processes range from high-energy physics interactions such as in $\bar p+p$ collisions at the Tevatron \cite{Grannis-Tevatron,Denisov-Tevatron} or $p+p$ collisions at the Large Hadron Collider (LHC) \cite{Dissertori:2012cra} to lower energy electron or hadron collisions on proton as well as nuclear targets at Jefferson Lab \cite{JLab10,Dudek:2012vr}. The former are designed to study QCD and electroweak interactions at the parton level, as well as to search for the Higgs boson and for physics beyond the standard model. The latter are designed to explore how QCD builds a hadron out of 3 valence quarks, how its quantum numbers are built up from these and from the ``sea'' of quark-antiquark pairs and gluons, how its properties change in a nuclear medium, and how a nucleus as a bound state of protons and neutrons emerges from the underlying microscopic quark and gluon degrees of freedom. 

One powerful set of tools that exploit the PDF universality are the so called ``global PDF fits'' \cite{Forte:2013wc,Jimenez-Delgado:2013sma,Rojo:2013fta}. Their original \emph{ra\^ison d'\^etre} is to utilize experimental data from a number of processes, combined with perturbative QCD calculations of the relevant partonic cross sections, in order to extract the non perturbatively calculable parton distributions. These in turn can be used to calculate processes not included in the fits, for example the expected rates of Higgs boson production in several channels, or various standard or beyond-the-standard model cross sections, such as production of $W'$ and $Z'$ gauge bosons, Kaluza-Klein resonances, gluinos, and so on. In this sense, increasing the number of data points to be fitted by including more processes, and improving the theoretical calculations to include a larger portion of the available kinematics, is extremely useful to reduce the uncertainties in the extracted PDFs, thus yielding more accurate theoretical predictions. This is one way in which lower energy hadronic and nuclear data, which typically access lower momentum scales and larger parton momentum fractions $x$ inside a proton than at colliders, can improve the study of high-energy processes.

However, as I will argue here, this phenomenologically very important connection is not necessarily the farthest reaching. Indeed, a novel aspect of large-$x$ global fits is their ability to
connect elements of high-energy physics with hadronic and nuclear
physics at medium energy \cite{Accardi:2013pra,Brady:2011hb}.  
For example, data
on $W$ and $Z$ boson production at forward rapidity at the Tevatron
and LHC, that extend to large values of $x$, can constrain the
extrapolation of PDFs to $x=1$, where these are sensitive to different
mechanisms of quark confinement \cite{Melnitchouk:1995fc, Holt:2010vj}.
Similarly, one can contrast these and other weak
interaction processes on \emph{proton} targets to DIS on \emph{deuterium} targets, which allow one to extract a nuclear interaction dependent $d$ quark distribution at large $x$ 
\cite{Accardi:2009br, Accardi:2011fa, Owens:2012bv}. Therefore, one can constrain in novel ways the nuclear dynamics and study the differences between bound and free protons, for which current models offer a large range of predictions.
Global fits can thus relate high-energy experiments at the Tevatron
 or LHC with nuclear physics experiments at lower energy facilities such as Jefferson Lab. I believe this interplay of particle and nuclear physics will prove very fruitful in the coming years, especially taking advantage of new weak interaction data from Jefferson Lab, RHIC and the LHC, and the planned high-energy Electron-Ion Collider in the US \cite{Accardi:2012qut,LRP2015}.

In this talk, I review the recent advances in global fits of PDFs of the proton as well as of the nucleus, illustrating some of the above mentioned connections. I will start by discussing proton PDF fits, that in fact utilize not only proton but also deuteron targets (and in the near future will also benefit from triton and Helium-3 targets, as well). As an interlude, I will discuss the potential dangers of utilizing heavier targets in these fits, specifically analyzing the case of dimuon production in neutrino-nucleus scattering. I will then conclude with a review of nuclear parton distribution function (nPDF) fits -- where the goal is to study the partonic structure of the nucleus rather than the free proton's -- and how these may (should!) be usefully combined with proton PDF fits to the benefit of both.

\section{Proton PDFs}  

\begin{figure}[tbh]
  \centering
  \includegraphics[width=0.80\linewidth]
                  {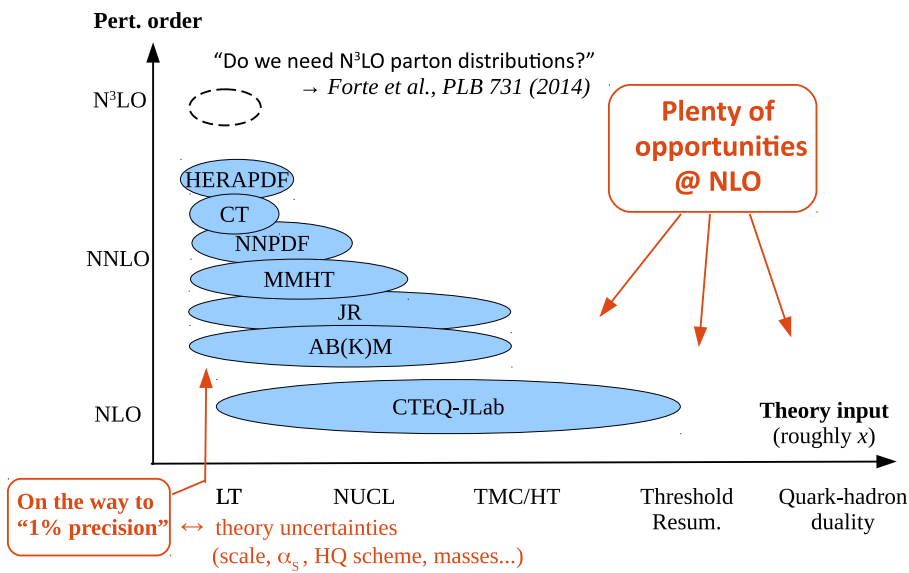}
  \caption{A PDF landscape. The major PDF fits are included in the horizontal balloons.}
  \label{fig:PDF-landscape}
\end{figure}

New fits have recently appeared on the arXiv, and a few have been announced or discussed at this conference \cite{Harland-Lang:2014zoa,Dulat:2015mca,DIS-Melnitchouk,Abramowicz:2015mha}.
These, together with the slightly older ones \cite{Ball:2014uwa,Jimenez-Delgado:2014twa,Alekhin:2013nda,Owens:2012bv}
can be usefully represented on the PDF landscape pictured in Figure~\ref{fig:PDF-landscape}, where the vertical axis represents the order of perturbative expansion considered in the calculation of the observables included in the fit, and the horizontal axis lists in rough order of increasing $x$ values a number of theoretical corrections to leading-twist calculations.

The bulk of LHC (and Tevatron) measurements is at mid-rapidity values, and therefore sensitive to small values of $x\approx (M/\sqrt{s})\exp(y)$, with $\sqrt s$ the center of mass energy, and $M$, $y$ the mass and rapidity of the observed final state. The often stated goal here is to reach ``1\% precision'' in the calculation of these observables, a large part of which can be achieved by improving the perturbative order in $\alpha_s$ at which these are made. This in turn requires PDF global fits utilizing perturbative calculations at the same order. The current state of the art, both in theory \cite{DIS-Forte,DIS-Siegert,DIS-Petriello} and global fitting efforts \cite{Rojo:2013fta}, is Next-to-Next-to-Leading order (NNLO). Calculations exist at this order for most of the observables included in global fits, and the missing ones will be completed soon enough. 
A few brave souls even started extending the perturbative calculations to Next-to-Next-to-Next-to-Leading Order level. (Given their complexity and the time they require to be completed, one can naturally ask ``Do we need N$^3$LO parton distributions?''; a possible answer can be found in Reference~\cite{Forte:2013mda}.) Nonetheless, one should note that for observables with more complicated final states, and indeed for many BSM signals, we still have to content ourselves with NLO calculations, which justifies the continued need for corresponding NLO fits. 

If one, however, is interested either in forward rapidity observables, or in large invariant mass final states, the parton's momentum fraction can easily exceed 0.1, see for example the recent CMS and ATLAS 3$\sigma$ ``bumps'' that are energizing the high-energy physics community \cite{Aad:2015owa}. Signals of BSM physics are however also sought as deviations from SM expectations in large invariant mass spectra even without a clearly identifiable bump, in which case one needs a much higher precision in the corresponding QCD calculation to claim discovery than available right now.
In this case increasing the precision of the extracted large-$x$ PDFs requires utilization of data in kinematic regions where a number of theoretical corrections beyond power-counting in the strong coupling constant. These include not only $1/Q^2$-suppressed corrections such as Target Mass Corrections (TMCs) and higher-twist (HT) corrections (that are avoided in typical PDF fits by requiring large enough cuts on the invariant mass of the final states, and thereby excluding data sensitive to large-$x$ PDFs), but also  non-power-suppressed corrections such as nuclear binding and Fermi motion in deuteron targets (necessary for up/down quark flavor separation) and threshold resummation (accounting for the reduced phase space for gluon radiation at large $x$, and affecting all observables). The theoretical uncertainties involved in these corrections are larger than the gain one can obtain by going to NNLO in the calculation, and a lot of theoretical and phenomenological progress can be made at NLO level before the need of going ``Next'' will be really felt.

In this regard, a lot of progress has been made in the last few years thanks in particular to the CTEQ-JLab (CJ) collaboration \cite{Accardi:2009br, Accardi:2011fa, Owens:2012bv}, that has performed a series of global PDF fits systematically including target mass effects (proportional to $M^2/Q^2$) and higher-twist corrections originating from multi-parton correlations in the nucleon (proportional to $\Lambda^2/Q^2$, with $\Lambda$ a typical hadronic scale of order 0.1-1 GeV), as well as nuclear effects in deuteron targets. Furthermore, a flexible $d$-quark parametrization was utilized in order to test what the data indicate for the behavior of $d/u$ ratio as $x \rightarrow 1$, which in many non perturbative models of the proton is expected to reach a value between 0 and 0.5 depending on the assumptions \cite{Roberts:2013mja}. Very importantly, a quantitative study of the associated theoretical uncertainties has been performed, see \cite{Accardi:2013pra} for a review of the results. 

The CJ studies -- that, in fairness, have built upon of earlier work by  Alekhin {\it et al.} \cite{Alekhin:2000ch,Alekhin:2012ig} and Martin {\it et al.} \cite{Martin:2003sk}) -- and the availability of new large-$x$ data from Jefferson Lab, Tevatron and LHC (with more to come in the near future from RHIC and the E906 experiment at Fermilab) have rekindled in the community the interest in large-$x$ PDF studies. Some of these corrections have then been individually considered in recent CT14 fits (flexible $d$-quark parametrization \cite{Dulat:2015mca}), NNPDF3.0 (TMCs \cite{Ball:2014uwa}), and MMHT14 (nuclear corrections \cite{Harland-Lang:2014zoa}), substantially confirming the CJ results. All these collaborations, however, still utilize large cuts on the invariant mass of DIS final states, thereby limiting their reach in parton's momentum fraction $x$ as compared to CJ and ABM.

Having brought TMCs, HT, and nuclear corrections under control, I believe the next frontier in large-$x$ PDF fits will be the consistent inclusion of threshold resummation effects on top of the mentioned corrections and a careful study of the interplay of all these effects, see for example \cite{Accardi:2014qda}. Resummation may also be important to relieve tensions between small Feynman $x$ vector boson data from the Fermilab's E866 experiment and other data in global fits \cite{Alekhin:2006zm}, and is necessary to utilize direct photon data, that would otherwise be fully incompatible with other observables, to improve up to 10\% the uncertainty in large-$x$ gluon PDFs \cite{Sato-PhD}.
A first fit including threshold resummation for a variety of observables (but unfortunately still excluding most of the large-$x$ DIS data in the threshold region by means of a large invariant mass cut) has been finalized after the end of this conference by the NNPDF collaboration \cite{Bonvini:2015ira}.
The time is ripe for all to follow their cue.

\section{Protons and deuterons: All for one and one for all}

We can now discuss with an example how global fits can tie together the realms of nuclear and high-energy physics. In particular, we will see how one can study the nuclear physics of the deuteron utilizing proton targets (!) and improve the proton's $d$ and $u$-quark PDF precision as a by-product.

\begin{figure}[b]
  \centering
  \includegraphics[width=0.48\linewidth]
                  {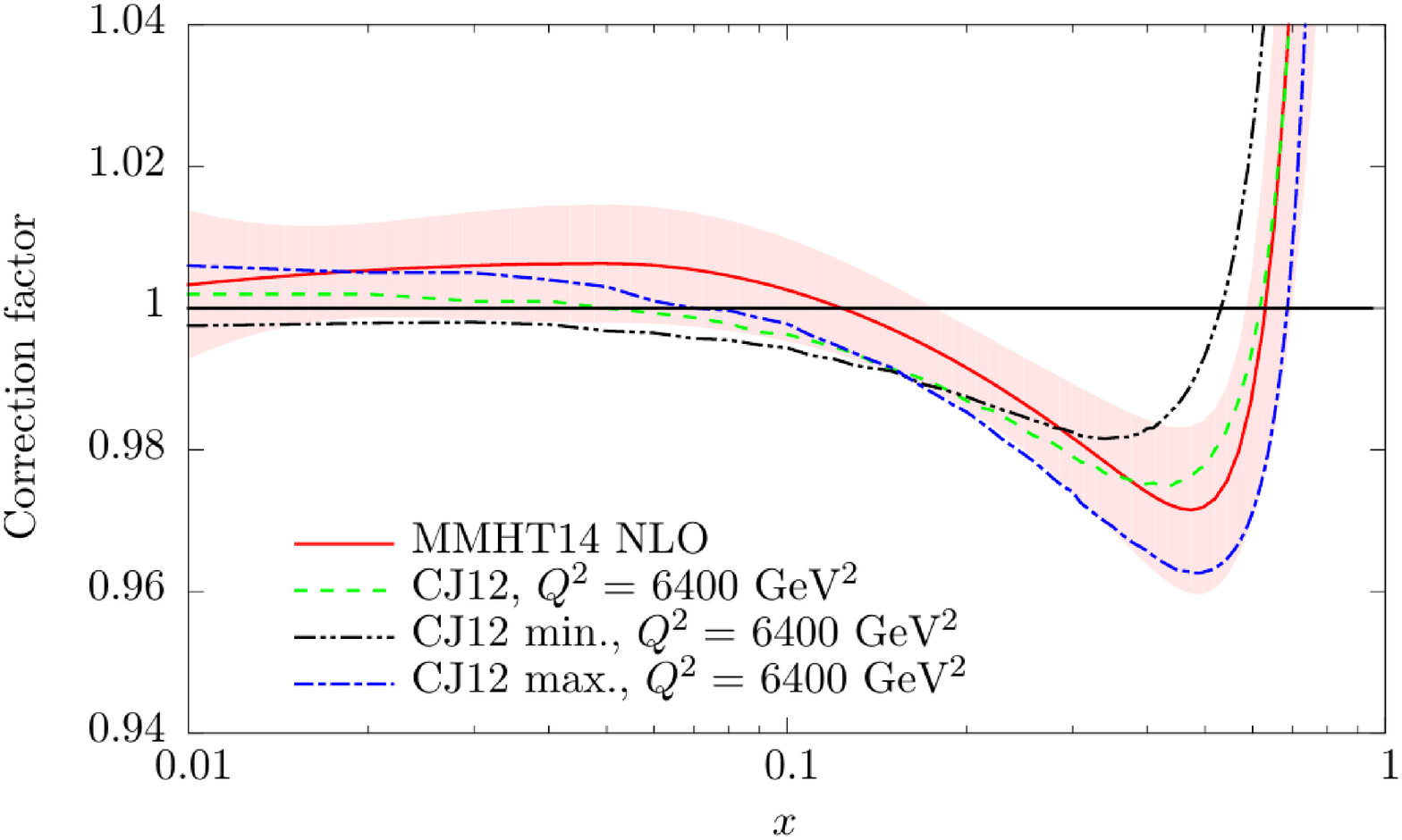}
  \includegraphics[width=0.48\linewidth
                  ,trim=0 200 20 0]
                  {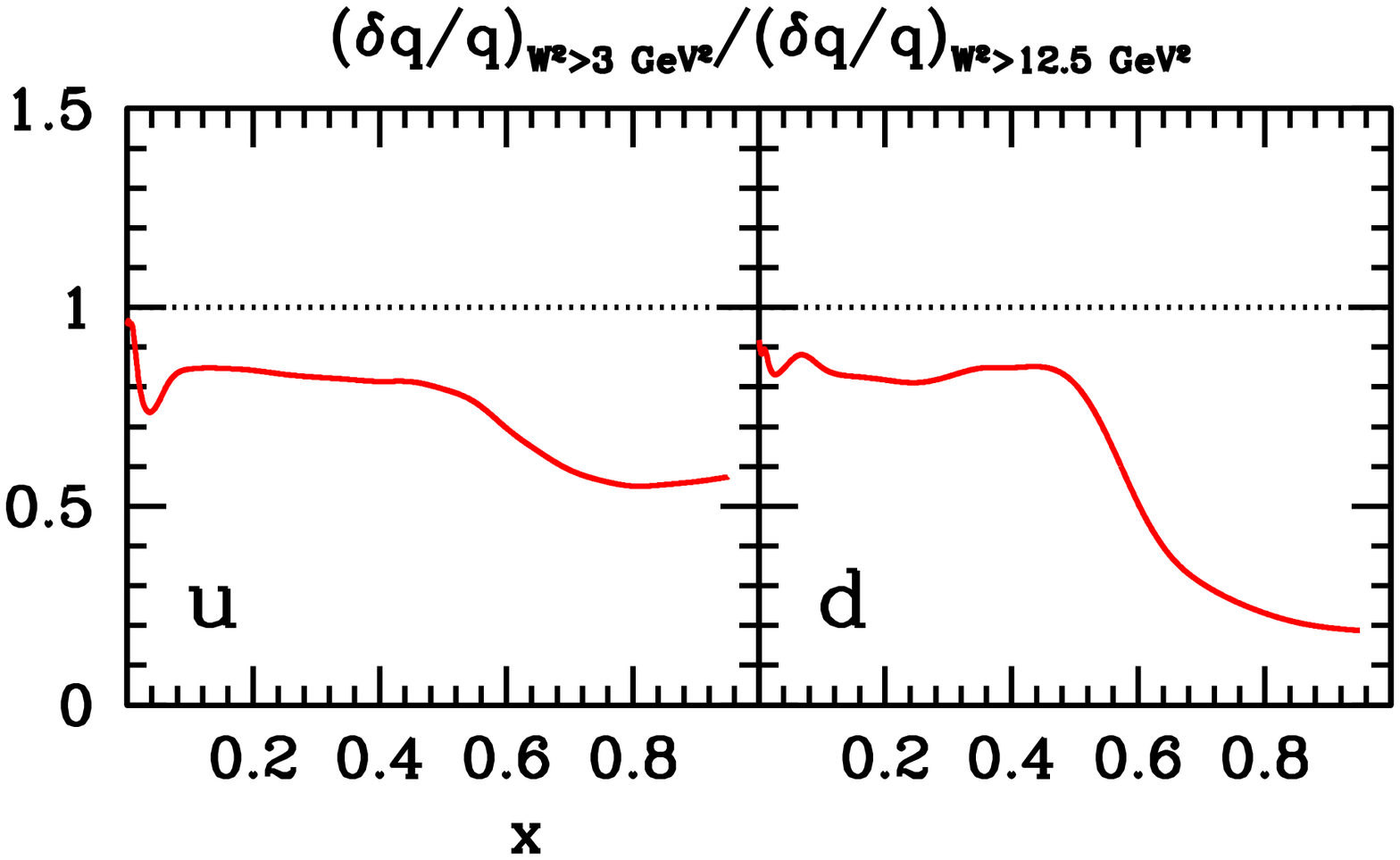}
  \caption{{\em Left:} The $F2(D)/F2(p+n)$ extracted in the CJ12 and MMHT14 global QCD fits. Plot from \cite{Harland-Lang:2014zoa}. {\em Right:} Relative uncertainty gain in the CJ12 $u$ and $d$ quarks when utilizing a $W^2>3$ GeV$^2$ cut on DIS data compared to a $W^2=12.5$ GeV$^2$ cut utilized in most other global PDF fits. Plot from \cite{Accardi:2013pra}}
  \label{fig:DN_and_du_ratios}
\end{figure}

As mentioned, parton distribution functions within a nucleus are different than in a superposition of isolated nucleons, and corrections for initial state nuclear effects such as binding, Fermi motion, and the offshellness of the nucleons are needed to utilize even light targets such as the deuteron in a global fit. However, there is no large scale to separate the nuclear interactions from the nucleon-level scattering, and calculation are largely model-dependent. There are 2 approaches to including nuclear corrections in global PDF fits: one is to parametrize the $F_2(D)/F_2(p+n)$ ratio and fit the needed parameters alongside the PDFs \cite{Harland-Lang:2014zoa};
alternatively, one can calculate the corrections using appropriate nuclear theory approximations, such as in the smearing approach where one convolutes the cross section for scattering on a bound nucleon with a suitable nuclear spectral function \cite{Owens:2012bv}. The advantage of the former method is its model independence (aside from, of course, the bias introduced by the choice of parametrization); but the nuclear physics output is very limited. In the second approach, one can compare the adopted nuclear modeling to data and verify or falsify the assumptions, thus gaining information on the nuclear dynamics of the targets.

This would be a nearly impossible goal if one was to only use DIS data: in the fit, the $d$ quark PDF would adapt and absorb the variations in nuclear models \cite{Accardi:2011fa}. 
However, global QCD fits can go beyond that, and compare high precision $d$-quark sensitive data from a variety of processes on proton targets with equally $d$-quark sensitive data from DIS on deuteron targets.
Given the high precision and large-$x$ reach of the recent D0 data on reconstructed $W$ charge asymmetry, one may expect that utilizing the ``wrong'' nuclear model will yield a $d$ quark trying to accommodate both data set, but with large $\chi^2$ for both. Therefore the fit will be able to select a reasonable set of nuclear corrections, and quantify the theoretical uncertainties inherent in the nuclear modeling. Conversely, with the right nuclear modeling in place, the much more abundant DIS data will reduce as much as possible the statistical uncertainties in the $d$ quark extraction.

\begin{figure}[tbh]
  \centering
  \includegraphics[width=0.50\linewidth
                  ,trim= 0 30 0 0]
                  {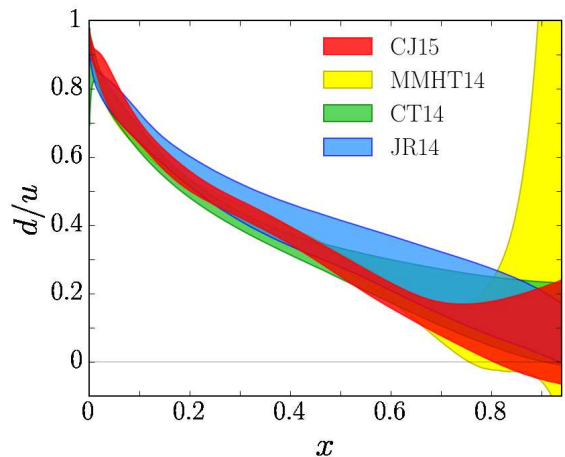}
  \caption{Large-$x$ $d/u$ ratios from various global fits \cite{DIS-Melnitchouk}.}
  \label{fig:du}
\end{figure}

This is precisely the message of the CJ12/CJ15 \cite{Owens:2012bv,DIS-Melnitchouk} and MMHT14 \cite{Harland-Lang:2014zoa} fits. Moreover, both agree in the size and shape of the obtained $F_2(D)/F_2(p+n)$ ratio (Figure~\ref{fig:DN_and_du_ratios} left), showing the complementarity of the two approaches to nuclear corrections. The statistical power of the extended DIS data set utilized in the CJ fits is shown in 
Figure~\ref{fig:DN_and_du_ratios} right. As a consequence one can, for example, extract a rather precise $d/u$ quark ratio, with controlled nuclear uncertainties, see Figure~\ref{fig:du}. The data constrain the ratio up to $x\approx 0.85$ and the extrapolation indicates $d/u(x=1) \lesssim 0.3$ thus excluding a few of the available non-perturbative calculations discussed, e.g., in Ref.~\cite{Roberts:2013mja}. Note that MMHT14 do not use a flexible $d$-quark parametrization so their $(d/u)_{MMHT14} \rightarrow 0,\infty$ only; CT14 do not include nuclear corrections and may be overestimating the ratio at large $x$. With precise data expected from the JLab 12 GeV upgrade (BONUS, Marathon, and Parity-Violating experiments), and in W asymmetry measurements at RHIC, these methods can be further extended to obtain yet more precise PDFs and nuclear correction constraints.

The increased precision of the extracted large-$x$ $u$- and $d$-quark PDFs can be finally leveraged to improve the precision in calculations of large rapidity and/or invariant mass observables at LHC for BSM physics searches, see for example Reference~\cite{Farry:2015xha}.

\section{Interlude: Strangeness and dimuon production in neutrino-nucleus scattering}

\begin{figure}[tbh]
  \centering
  \includegraphics[width=0.50\linewidth]
                  {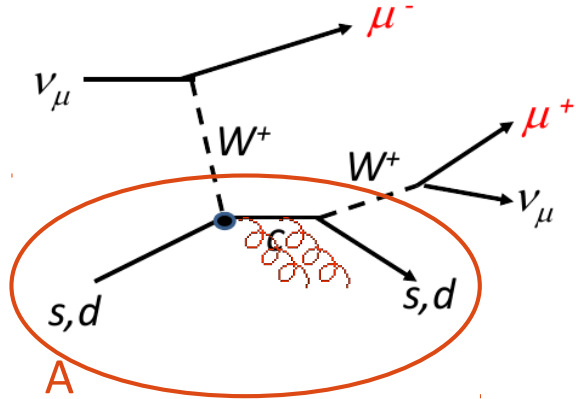}
  \caption{Final state interactions of the charm quark in dimuon production in $\nu+A$ collisions.}
  \label{fig:dimuons}
\end{figure}

Without using LHC data, most of the information on strange quark distributions comes from dimuon production in neutrino-nucleus scattering. In these reactions one can tag a neutrino scattering on a strange (or down) quark by detecting a dimuon pair in coincidence, see Figure~\ref{fig:dimuons}. This process is widely used in PDF fits to provide constraints on the strange quark that would otherwise be very weak. However, the presence of a nuclear target requires control of nuclear corrections. These come into 2 varieties: initial state nuclear modifications of the PDFs themselves, and final state interactions of the charm quark (and possibly of the D meson it mostly hadronizes into) as this traverse the nuclear target after the hard scattering.

On the one hand, initial state effects are partly under control using information coming from ``nuclear PDF'' fits, see next Section; however, these typically fit data on nucleus to deuteron ratios, treating the deuteron itself as a pair of unbound proton and neutrons calculated using the independently fitted nucleon PDFs themselves. This implies some double counting systematics, which is difficult to estimate unless a combined fit to proton and nuclear data is performed. Nonetheless the precision of the data themselves was considered not high enough for this to be a practical problem. However, this has recently changed with the recent availability of precise NOMAD and CHORUS data.

On the other hand, final state effects such as medium-induced gluon bremsstrahlung off the scattered charm quark, or nuclear absorption of the D meson (should this hadronize still inside the nucleus) are much harder to quantify, and little phenomenology exists for DIS on nuclear targets \cite{Accardi:2009qv}. For example, final state suppression of charm production has been predicted theoretically and observed in heavy ion production at RHIC (where the traversed medium is a hot Quark-Gluon Plasma) but with a much larger size than perturbative calculations had predicted, originating the so-called ``heavy quark puzzle'' \cite{Djordjevic:2014hka}; this casts doubts on how well one can even only estimate the size of the effect in a cold nuclear target.

Given the lack of theoretical and phenomenological control over final state in-medium suppression of the charm quark and its potentially large magnitude, it is dangerous to utilize dimuon data to obtain constraints on the proton strange quark PDF, and one risks to underestimate this by an uncontrolled amount. This is in fact what seems to be happening when comparing the strange quark extracted from LHC data on $W + c$ productions, where one observes a ratio $k= 2 s / (\bar u + \bar d) \approx 1$ compared to $k \approx 0.4$ obtained in fits utilizing only dimuon data; see \cite{Alekhin:2014sya} for a more detailed discussion and for an alternative interpretation of this mismatch. Indications of issues with strange quarks also come from semi-inclusive DIS kaon cross sections from the HERMES collaboration \cite{Airapetian:2008qf}, that differ in shape from those calculated using PDFs determined using dimuon data.

A measure of control -- or at least a check on the size of final state effects -- can be obtained by utilizing global PDF fits in analogy to what I discussed in the case of large-$x$ fits of $d$ quarks in the previous section.
Namely, global fits allow one to obtain constraints on nuclear effects on a given quark whenever data on proton targets sensitive to the same initial state parton flavors are available. This is precisely what can be done now thanks to the recent $p+p \rightarrow W,Z + c$ measurements from the LHC: by comparing these to dimuon production in $\nu+A$ DIS, one can {\it extract} final state effects from the nuclear data. This is a novel, phenomenologically important, and extremely exciting opportunity that opens up as soon as one is willing to combine nuclear and high-energy physics! 

When final state systematic effects are brought under control, one will then be able to utilize the full statistical power of the combined proton and nuclear targets data set to obtain the most precise strange quark fit possible. At this level of precision one will likely need to also control initial state effects, which is possible, in principle, within the framework of nuclear PDF fits to which we now turn our attention.

\begin{figure}[b]
  \centering
  \includegraphics[width=0.70\linewidth]
                  {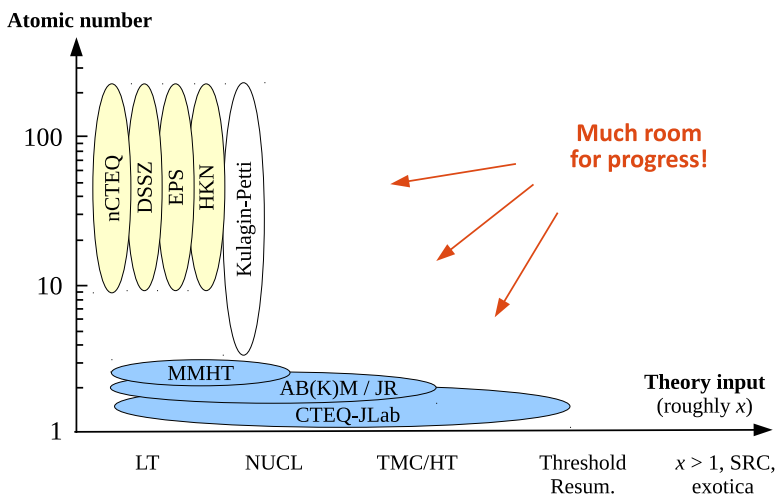}
  \caption{A nuclear PDF landscape. Recent nuclear PDF fits are included in the yellow vertical balloons, and proton PDF fits that treat deuteron targets with a measure of nuclear theory input are listed in the horizontal blue balloons. An analysis by Kulagin and Petti of nuclear DIS structure function utilizing a comprehensive modeling of nuclear effects (but without PDF fitting) is included in the empty balloon.}
  \label{fig:nPDF-landscape}
\end{figure}

\section{Nuclear PDFs}

In terms of elementary degrees of freedom, nuclei are made of quarks and gluons as much as protons are. It is therefore natural, and phenomenologically important for the description of hard scattering processes involving nuclear targets, to measure the so-called ``nuclear Parton Distribution Functions'' (nPDFs). This endeavor is immediately complicated by the existence of an extra variable, the atomic mass number $A$, as depicted in the nPDF landscape of Figure~\ref{fig:nPDF-landscape} left. At the same time, given the much reduced availability of data on nuclear targets both in number of data points and in kinematic coverage compared to the proton case, the need for precision in perturbative calculations is much reduced compared to proton PDF studies, and nPDF fits have been limited to NLO order so far. Therefore we can still still distribute nuclear PDFs fits and studies in a 2-dimensional plane.

Although QCD factorization and universality theorems for nuclear targets are not as firmly established as for proton targets due to both initial and final state interactions \cite{Qiu:1996zu,Kang:2013ufa}, nPDF fits assuming factorization and universality have proven to be phenomenologically quite successful in describing hard scattering processes in electron-, proton- and nucleus-nucleus collisions in a range of energies spanning the GeV and the TeV
\cite{Eskola:2009uj,Hirai:2007sx,deFlorian:2011fp,Kovarik:2015cma}.

\begin{figure}[tb]
  \centering
  \includegraphics[width=0.70\linewidth]
                  {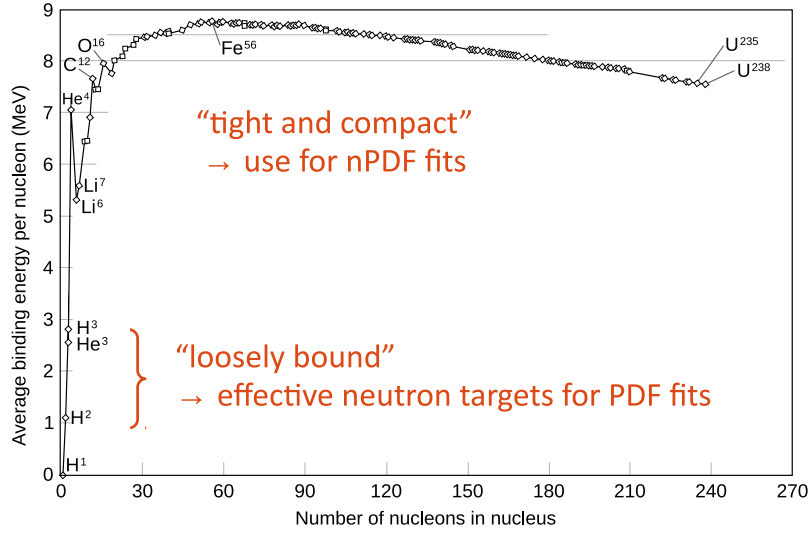}
  \caption{A chart of per-nucleon binding energies.}
  \label{fig:nuclear-chart}
\end{figure}

Looking at a chart of nuclear binding energies, such as that of Figure~\ref{fig:nuclear-chart}, it is clear that nuclei can be divided into two classes: tight and compact ``heavy nuclei'' with $A \gtrsim 12$ (including, however, also Helium-4 as an interesting exception), and the weakly bound $A\leq 3$ nuclei. The latter are weakly bound, and the effects of their nuclear interactions on hard scattering events are amenable to a controlled theoretical description. Thus they can be used as effective neutron targets for PDF fits, as discussed for the case of the deuteron in the previous sections. The former nuclei are more strongly bound, and the nuclear deformation of their parton distributions are the subject of nPDF fits. Intermediate between these, are the so-called ``light'' ions, that exhibit characteristics of both classes and not necessarily can be extrapolated from the behavior of either, as has become increasingly evident thanks in particular to recent high precision measurements at Jefferson Lab \cite{Malace:2014uea,Kulagin:2010gd,Seely:2009gt,Arrington:2012ax}. Much more data on DIS on light and heavy nuclei is expected from Jefferson Lab's 12 GeV upgrade \cite{JLab10,Dudek:2012vr}, as well as from the Electron-Ion Collider project that has just been endorsed in the NSAC long range plan released in October \cite{JLab10,Dudek:2012vr}.

Nuclear PDF fits divide in two classes depending on what quantity is parametrized and fitted. Traditionally \cite{Eskola:2009uj,Hirai:2007sx,deFlorian:2011fp}, nuclear PDFs are obtained using a multiplicative correction factor
\begin{align}
  R(x,Q_0)=f_k^A(x,Q_0,A)/f_k(x,Q_0) \ ,
\end{align}
where $f_k(x,Q)$ is the proton PDF for a parton of flavor $k$, and $f_k^A(x,Q,A)$ is the corresponding nPDF in a nucleus of atomic number $A$ (a functional dependence on $Z$ is possible, but typically neglected when considering heavy nuclei; it may, however, become important for light nuclei). The correction factor $R$ is parametrized -- in different ways by the different groups -- at an initial scale $Q_0$, and fitted to a variety of data, from DIS to dilepton and pion production in proton-nucleus scatterings. Alternatively, the nCTEQ collaboration \cite{Kovarik:2015cma} has introduced a ``native PDF parametrization'' in which the nPDFs themselves are anchored to an existing proton PDF parametrization, but have $A$-dependent parameters, {\em e.g.}, 
\begin{align}
  x f_k(x,Q_0,A) = c_0 x^{c_1}(1-x)^{c_2} e^{c_3 x}(1+e^{c_4}x)^{c_5}  
\end{align}
with
\begin{align}
  c_i = c_i(A) = c_{i,0} + c_{i,1}(1-A^{-c_{i,2}}) \ .
\end{align}
QCD evolution is then applied directly to the nPDFs, instead of assuming a $Q$-independent multiplicative ratio.

\begin{figure}[tb]
\vspace*{.2cm} \includegraphics[width=0.48\textwidth]{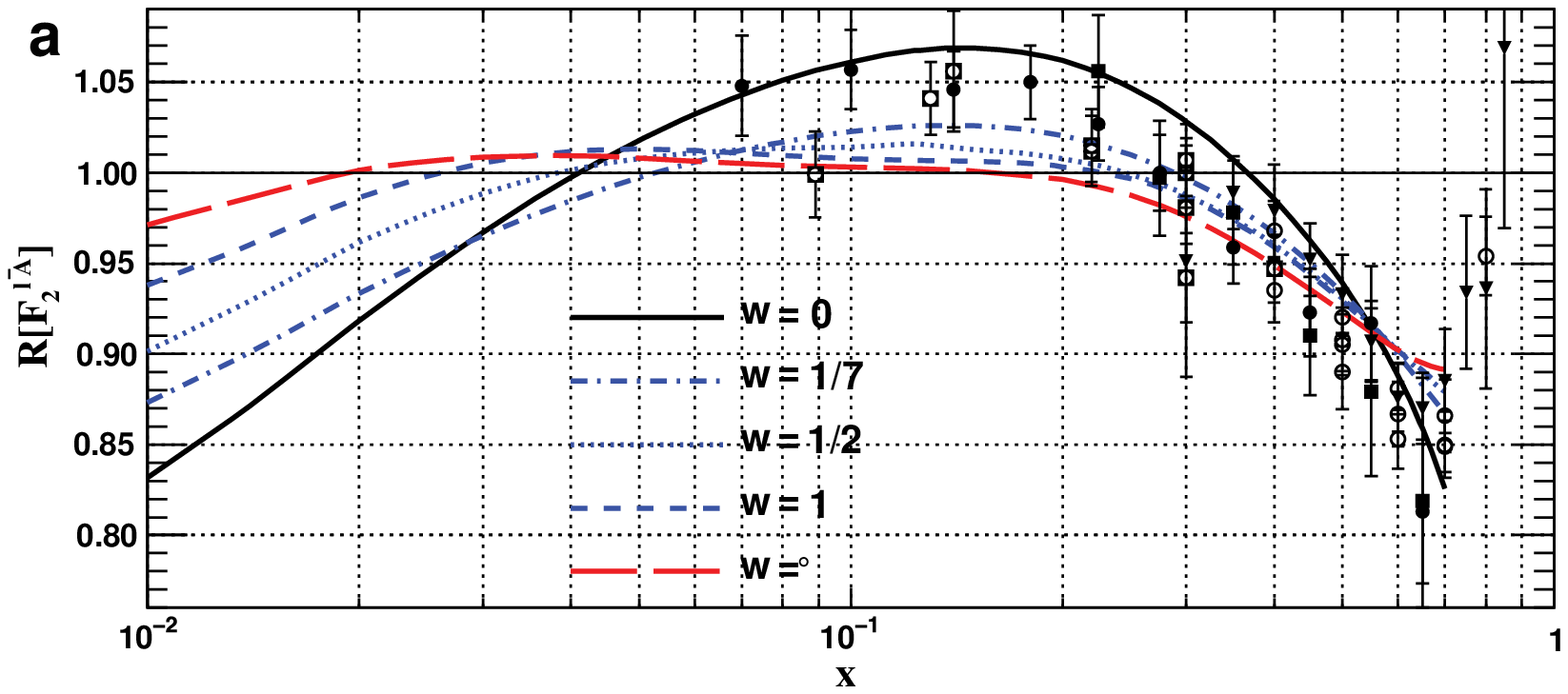}\hspace{0.4cm}
  \includegraphics[width=0.48\textwidth]{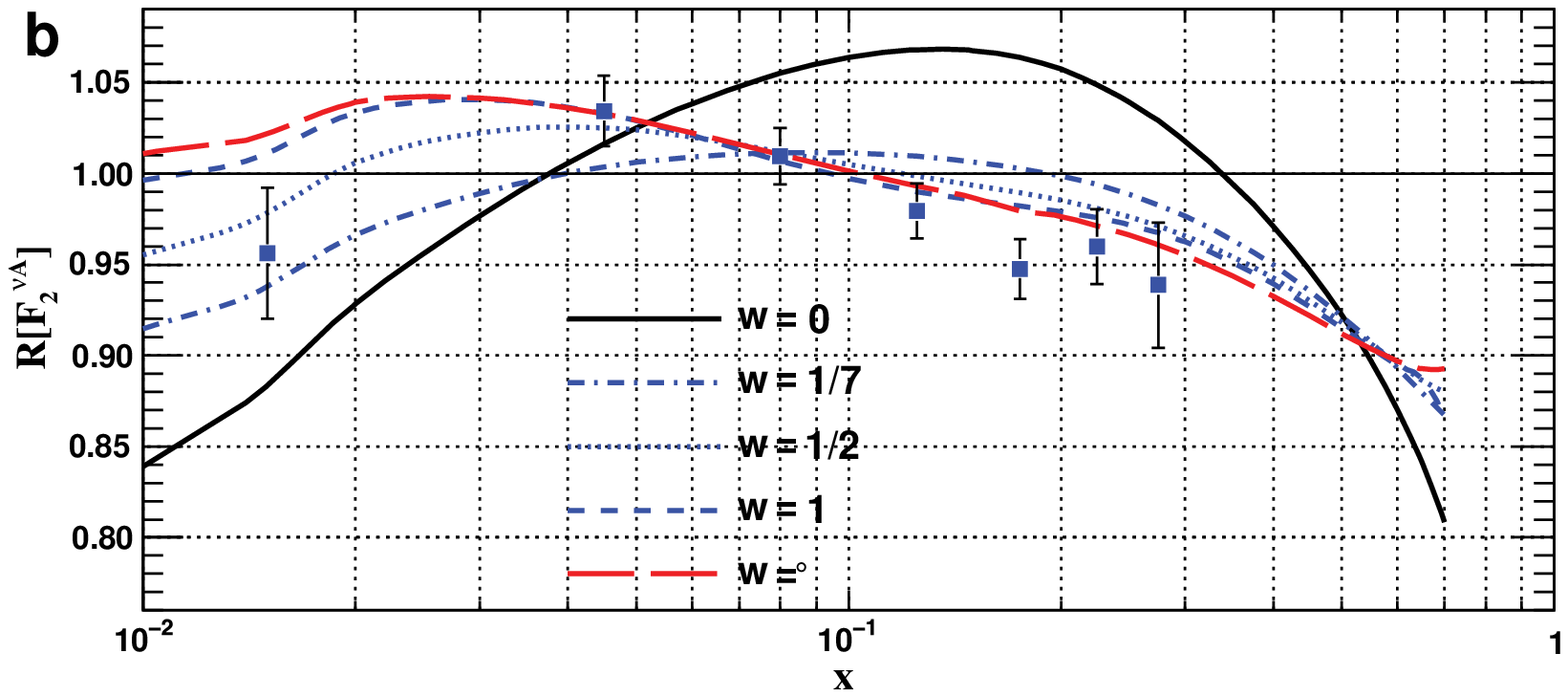}
  \caption{Ratios of Iron to deuteron $F_2$ structure functions in neutral vs. charged current DIS. Calculations using nCTEQ nPDFs are compared to experimental data. Plots taken from \cite{Kovarik:2010uv}.}.
  \label{fig:F2compare}
\end{figure}

An outstanding, and still unresolved, issue in nuclear DIS was highlighted in 2011 by the nCTEQ collaboration, who observed that it is not possible to reproduce the data in neutrino-nucleus scattering using nPDFs fitted to lepton-nucleus data \cite{Kovarik:2010uv}, see Figure~\ref{fig:F2compare}. The EPS fit authors, however, do not observe this tension between neutral and charged current scattering \cite{Paukkunen:2013grz,Paukkunen:2014nqa} (but do not take into account correlated errors in the NuTeV data in their analysys). The ensuing debate has not yet been settled. The question is nonetheless very important, because confirming the tension would point at deficiencies in our current theoretical understanding of deep inelastic lepton-nucleus scattering; in turn this would limit the information that can be obtained from, say, neutrino oscillation experiments \cite{deGouvea:2013onf}. New data from NOMAD and MINERVA, and many neutrino facilities to come on line in the near future will hopefully shed light on this important issue.

Current nPDF fits typically have high energy applications in mind \cite{Paukkunen:2014nqa}, and naively treat nuclei as a collection of $Z$ unbound proton and $A-Z$ neutron targets in regard to the treatment of the kinematics, and assume universality of the nuclear parton distributions. However, as the observed ``non-scaling'' behavior of light nuclei at JLab and the discussion on neutral vs. charged current data highlight, this may not be the case. It would be useful to infuse a measure of more realistic nuclear physics in nPDF fits, in order to make these more precise, and enlarge their phenomenological reach in an analogous way that, say, proton QCD fits are now becoming able to explore the nuclear physics of the deuteron. For example, one may want to: 
\begin{itemize}
\item 
enlarge the kinematic reach to higher $x$ and lower $Q^2$ values, that necessitates TMCs, HT, Fermi motion and binding corrections;
\item
study the region of  ``superfast quarks'' at $x>1$ that JLab 12 (and EIC) will open up \cite{Sargsian:2002wc};
\item
explore the nature of the EMC effect, and its connection to Short-Range-Correlations between protons an neutrons in a nucleus \cite{Arrington:2011xs};
\item
study the behavior of light nuclei by comparing these to protons and deuterons ($A \rightarrow 1$)  as treated in PDF fits, on the one hand, and to heavy nuclei ($A \gg 1$) on the other;
\item
study final-state charm quark propagation in cold nuclear matter as discussed in the previous Section, and, more in general, in-medium parton propagation and fragmentation \cite{Accardi:2009qv}.
\end{itemize}
The use of a ``native nuclear PDF'' framework is in my opinion the most natural choice for such a generalization, and I think that ``anchoring'' this to a nuclear-theory-infused PDF fit such as from the CTEQ-JLab collaboration would go a long way toward these goals.

\section{Concluding remarks}

In reviewing recent progress in parton distribution studies from protons to nuclei, I have highlighted how global QCD fits can be utilized as a mean to bring together the High-Energy Physics and the Hadronic and Nuclear Physics communities in a common program to study the partonic structure of nucleons and nuclei, and the QCD dynamics that governs hadrons and describes the emergence of nuclei from quark and gluon degrees of freedom. It also makes it possible to combine data from diverse experimental sources and obtain increased precision in PDF fitting over a larger kinematic range than it would be possible when keeping nuclear and free proton data separate.

I think this combined approach is very rich in theoretical consequences and phenomenological applications, encompassing studies of Beyond-the-Standard Model physics at the LHC, neutrino physics studies (where nuclear targets play a dominant role), and the understanding of the non-perturbative QCD dynamics governing partonic interactions within hadrons and nuclei.

Let's go for it!

\vskip.6cm\noindent
{\bf Acknowledgments:} 
This work was supported by the DOE contract No.~DE-AC05-06OR23177,
under which Jefferson Science Associates, LLC operates Jefferson Lab, and by the DOE contract No. DE-SC008791. I would also like to thank the organizers of DIS2015 for their warm hospitality, and for their patience while I prepared this contribution to the conference proceedings.

\end{document}